\documentclass[pdflatex,sn-aps]{sn-jnl}

\usepackage{graphicx}
\usepackage{pgfplots}
\usepackage{amsmath,amssymb,amsfonts}
\usepackage{booktabs}
\usepackage{multirow}
\usepackage{subcaption}
\usepackage{float} 

\begin{document}

\title[Scattering Transformer]{Scattering Transformer: A Training-Free Transformer Architecture for Heart Murmur Detection}

\author*[1]{\fnm{Rami} \sur{Zewail}}
\email{rami.zewail@ejust.edu.eg}

\affil*[1]{\orgdiv{Computer Science \& Engineering Department}, \orgaddress{\city{Alexandria}, \country{Egypt}}}

\abstract{In an attempt to address the need for skilled clinicians in heart sound interpretation, recent research efforts on automating cardiac auscultation have explored deep learning approaches. The majority of these approaches have been based on supervised learning that is always challenged in occasions where training data is limited. More recently, there has been a growing interest in potentials of pre-trained self-supervised audio foundation models for biomedical end tasks. Despite exhibiting promising results, these foundational models are typically computationally intensive. Within the context of automatic cardiac auscultation, this study explores a lightweight alternative to these general-purpose audio foundation models by introducing the Scattering Transformer, a novel, training-free  transformer architecture for heart murmur detection. The proposed method leverages standard wavelet scattering networks by introducing contextual dependencies in a transformer-like architecture without any backpropagation. We evaluate our approach on the public CirCor DigiScope dataset, directly comparing it against leading general-purpose foundational models. The Scattering Transformer achieves a Weighted Accuracy(WAR) of 0.786 and an Unweighted Average Recall(UAR) of 0.697, demonstrating performance highly competitive with contemporary state of the art methods. This study establishes the Scattering Transformer as a viable and promising alternative in resource-constrained setups.}


\maketitle

\section{Introduction}\label{sec:introduction}
Cardiovascular diseases (CVDs) are the number one cause of death in the world, killing about one-third of all global deaths \cite{who_cvd}. Listening to heart sounds with a stethoscope, also known as cardiac auscultation, is considered the most common and least expensive way to do a preliminary cardiac screening. Nevertheless, the diagnostic accuracy of this technique is highly appendant upon the clinician's experience and proficiency. Hence, it frequently results in considerable inter-observer discrepancies. This is especially true in case of shortage in expert cardiologists \cite{ren2024comprehensive, niizumi2024exploring, panah2023exploring}. 
To address the challenges in manual auscultation, there has been a growing interest in the potentials of artificial intelligence in automated analysis of digital phonocardiogram (PCG) signals. Earlier research has focused on conventional machine learning models integrated with engineered features \cite{dwivedi2019algorithms, kong2020panns, gong2021ast}. Recent research directions have explored deep learning techniques. Convolutional Neural Networks (CNNs) and Recurrent Neural Networks (RNNs) have shown a lot of promise in classifying heart sounds by learning features from time-frequency representations of the audio \cite{niizumi2023masked}. However, these supervised DL methods usually need a lot of carefully labeled data, which can be scarce in the medical field. In general, despite being successful, deep learning methods often need a lot of data. This becomes problematic in domain-specific applications such as heart sound datasets. Moreover, there aren't many pre-trained models that are relevant to the field \cite{niizumi2024exploring}. More recently, the research landscape has shifted toward self-supervised learning (SSL) and the utilization of large-scale, pre-trained audio foundation models to reduce dependence on labeled data \cite{niizumi2024exploring}. The majority of these models adopt a transformer architecture with heavly reliance on attention mechanisms. When fine-tuned for biomedical tasks like heart murmur detection \cite{oliveira2022circor, niizumi2021byol}, models like AST \cite{gong2021ast}, wav2vec 2.0 \cite{baevski2020wav2vec}, and M2D \cite{niizumi2023masked} have shown state-of-the-art performance. These models are often pre-trained on large, general-purpose audio datasets like AudioSet \cite{gemmeke2017audio}. These foundational models are very accurate, but they are also very big and require a lot of computational power, which makes them hard to use on low-power edge devices or in places with limited computational infrastructure. This creates a research gap: we need a general-purpose biomedical signal representation that is both data and energy efficient, without losing performance. Within this context, we present the Scattering Transformer, an innovative lightweight transformer architecture that is fundamentally training-free. The suggested model leverages upon the appealing properties of wavelet scattering networks and adds a parameter-free transformer architecture to enhance its contextualization ability. We directly compare our proposed method to the best large-scale audio foundation models to see if it could be a useful and competitive way to find heart murmurs in the public CirCor DigiScope dataset from the 2022 George B. Moody PhysioNet Challenge \cite{reyna2023heart, koike2020audio}. 

The main contributions of this work are:
\begin{itemize}
    \item \textbf{A  training-free transformer architecture based on wavelet scattering networks:} We present the Scattering Transformer, that extends standard wavelet scattering networks by incorporating attention and position encoding in a transformer-like architecture.
    \item \textbf{Competitive performance against foundational models:} Within the application of bio-signal representation, we show that our lightweight Scattering Transformer exhibits competitive performance against the state of the art pre-trained transformer-based models that were trained on general audio.
    \item \textbf{A baseline for a lightweight transformer-like architecture for efficient biomedical Signal representation:} Our research establishes a blueprint for contextualizing standard deep scattering representation using a training-free transformer architecture. The result is a resource-aware representation that is competitive with state-of-the-art pre-trained audio foundational models. The proposed scattering transformer is exemplified in biomedical signal representation and a terminal task of heart murmur detection.
\end{itemize}

The rest of the paper is organized as follows: the related work is presented in section \ref{sec:related_work}. The details of the proposed scattering transformer are presented in section \ref{sec:methodology}. Experiments and results are detailed in section \ref{sec:experiments}. Discussion is presented in section \ref{sec:discussion}. Finally, a conclusion is presented in section \ref{sec:conclusion}.

\section{Related work}\label{sec:related_work}
This section provides a survey of the most relevant work in the literature addressing challenges related to automated cardiac auscultation and heart murmur detection.

\subsection{The Heart Murmur Detection Research Landscape}
Driven by the increasing integration of AI into cardiac diagnostics, research efforts such as the George B. Moody PhysioNet Challenge have been established by the IEEE \cite{reyna2023heart}. The 2022 PhysioNet Challenge, in particular, focused on heart murmur detection, introducing a core problem of identifying murmurs from multi-location phonocardiogram (PCG) recordings using only training samples \cite{reyna2023heart}. This has motivated the development of a variety of deep learning-based heart murmur detection approaches \cite{dwivedi2019algorithms, ren2024comprehensive}. More recent work in the literature has broadened to include transfer learning from large audio datasets \cite{niizumi2024exploring, koike2020audio}, the adaptation of transformer architectures \cite{Han2025ENACTHeart, gong2021ast}, and self-supervised learning (SSL) paradigms \cite{panah2023exploring, niizumi2021byol}.
\subsection{Large-Scale Pre-trained and Foundation Models}
The challenge of limited annotated medical data has made transfer learning a highly relevant strategy. One prominent approach involves fine-tuning general-purpose audio models pre-trained on massive, out-of-domain datasets like AudioSet \cite{gemmeke2017audio}. A systematic study by Niizumi et al. \cite{niizumi2024exploring} provides a crucial benchmark, highlighting that transformer-based architectures such as the Audio Spectrogram Transformer (AST) \cite{gong2021ast} show a distinct advantage over CNNs for this task. Their work also established the state-of-the-art Masked Modeling Duo (M2D) \cite{niizumi2023masked}, an SSL-trained transformer, which achieved new performance benchmarks by effectively transferring representations learned from over two million general audio clips to the murmur detection task. An alternative strategy focuses on learning representations directly from in-domain data. Panah et al. \cite{panah2023exploring} pioneered this for heart sounds by adapting the wav2vec 2.0 model \cite{baevski2020wav2vec}, a self-supervised transformer from the speech domain. By finetuning the model on unlabeled heart sounds from the CirCor DigiScope dataset \cite{oliveira2022circor}, they achieved competitive performance maintaining high accuracy even when fine-tuning with as little as 13\% of the available labeled data. While these approaches are powerful, they both rely on computationally expensive, data-driven learning to optimize millions of parameters and have a large memory footprint. Our work directly contrasts with this by introducing a training-free transformer-like architecture based on wavelet scattering networks \cite{bruna2013invariant}.

\subsection{Hybrid Architectures and Data-Efficient Methods}
Recognizing that a single architecture may not capture all signal complexities, many recent studies have proposed hybrid and ensemble models. For heart sound analysis, examples include ENACT-Heart, which ensembles a Vision Transformer (ViT) and a CNN \cite{Han2025ENACTHeart}, and HSDFE-Net, a dual-branch network that processes conventional and bi-spectrum features in parallel \cite{hsdfenet}. These methods validate the principle of combining complementary components. The Scattering Transformer extends this hybrid approach but innovates by fusing a non-trainable lightweight deep wavelet scattering network with a non-trainable contextualizer architecture, maximizing efficiency.

\subsection{Positioning of the Proposed Method}
This section describes how the proposed framework is positioned into the corpus of relevant literature. The core novelty lies in the integration of an inherently lightweight wavelet scattering front-end with a training-free contextualization module. In contrast to the works of Panah et al. \cite{panah2023exploring} and Niizumi et al. \cite{niizumi2024exploring, niizumi2023masked}, avoids the need to train or fine-tune large models on massive datasets. Moreover, unlike the transfer learning-based methods presented in other studies \cite{niizumi2024exploring, koike2020audio}, our proposed method employs attention context-aware wavelet scattering-based embeddings at its core. Finally, the method presented in this work builds upon the direction in the literature tackling challenges related to data and computational efficiency \cite{ren2024comprehensive}.
\section{Methodology}\label{sec:methodology}
\begin{figure}[H] 
    \centering
    \begin{subfigure}[b]{\textwidth}
        \centering
        \includegraphics[width=\linewidth]{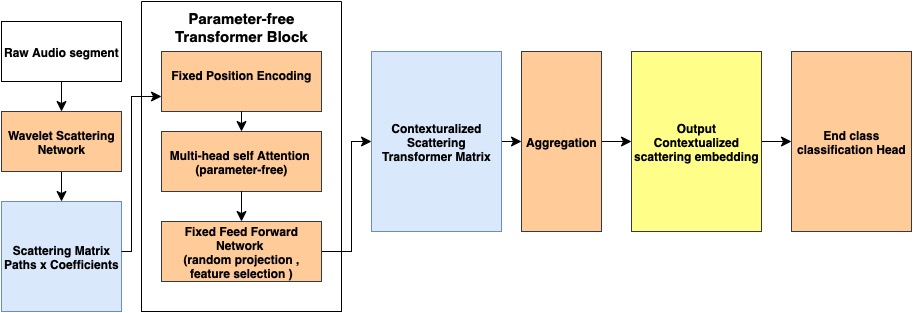}
        \caption{Paths-as-Sequence Mode: The contextualization module operates on the scattering matrix from a single audio segment, capturing dependencies between different scattering paths.}
        \label{fig:mode1}
    \end{subfigure}
    \vfill
    \begin{subfigure}[b]{\textwidth}
        \centering
        \includegraphics[width=\linewidth]{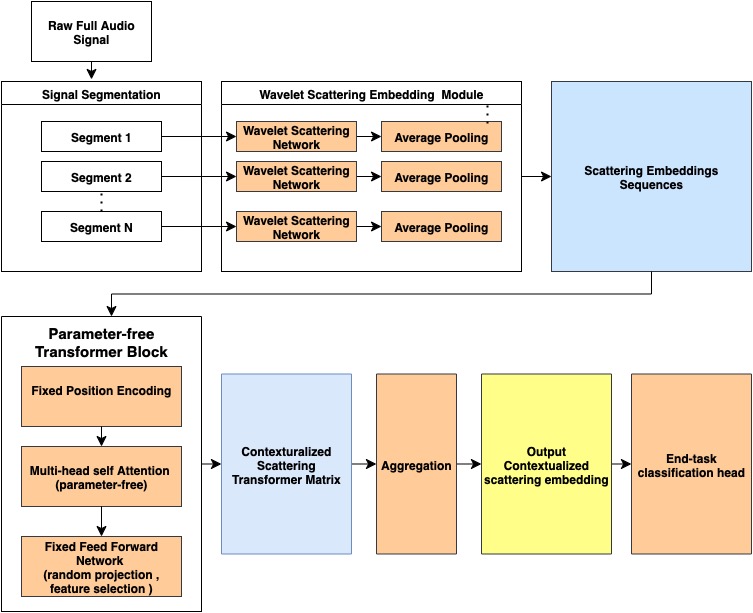}
        \caption{Multi-Segment Mode: The audio signal is segmented, and a scattering embedding is generated for each segment. The contextualization module then captures dependencies across these temporal segments.}
        \label{fig:mode2}
    \end{subfigure}
    \caption{The two operational modes of the proposed Scattering Transformer architecture. Both modes utilize the same core parameter-free transformer block but differ in how the input sequence is constructed.}
    \label{fig:overview}
\end{figure}

This section presents the details of the proposed Scattering Transformer that incorporates position encoding and self-attention to the wavelet scattering networks. A top-level overview of the proposed architecture is given in Figure \ref{fig:overview}.The proposed method is evaluated for an end task of supervised classification for detection of murmur heart conditions.

\subsection{Wavelet Scattering Network Front-End}
The front end of the proposed Scattering Transformer is a standard wavelet scattering network (WSN) block. A wavelet scattering network is a deep architecture that efficiently extracts low-variance deep signal representation \cite{bruna2013invariant}. Unlike traditional CNNs, a standard WSN utilizes predefined wavelet and scaling filters, eliminating the need for data-driven filter learning. The wavelet scattering network is constructed by iteratively applying these three operations: convolution with a wavelet filter bank, a non-linear modulus operator, and an average pooling operation. The m-th order scattering coefficients are defined as in Equation \ref{eq:scattering_coeff}.
\begin{equation}
    S_m f = |\ldots|f * \psi_{\lambda_1}| * \ldots * \psi_{\lambda_m}| * \phi_J|
    \label{eq:scattering_coeff}
\end{equation}
Where $f$ is the input biosignal, $\psi_{\lambda_k}$ is the wavelet filter at scale $\lambda_k$, $*$ is the convolution operator, $|.|$ is the modulus operator, and $\phi_J$ is the low-pass scaling function that introduces invariance at scale $J$. The final scattering feature matrix is obtained by aggregating the scattering coefficients from all orders up to a maximal order M, as given in Equation \ref{eq:scattering_matrix}.
\begin{equation}
    Sf = \{S_m f\}_{0 \le m \le M}
    \label{eq:scattering_matrix}
\end{equation}
The resulting scattering matrix is then used as an input for the subsequent contextualization stage.
\subsection{Training-free Contextualization}
While the wavelet scattering network produces an efficient deep representation, it does not explicitly model the long-range dependencies within the signal. To address this, the scattering features are passed to a parameter-free contextualization block, which functions as a fixed contextualizer. This block captures the global relationships across the feature sequence without any trainable weights.
\subsubsection{Positional Encoding}
A standard sinusoidal positional encoding, PE, is added to the sequence of scattering features, $X_{scat}$.
\begin{equation}
    X_{pos} = X_{scat} + PE
    \label{eq:pos_encoding}
\end{equation}
The encoding for position $\text{pos}$ and dimension $i$ is calculated as shown in Equations \ref{eq:pe_sin} and \ref{eq:pe_cos}.
\begin{equation}
    PE_{(pos, 2i)} = \sin(\text{pos} / 10000^{2i/d_{\text{model}}})
    \label{eq:pe_sin}
\end{equation}
\begin{equation}
    PE_{(pos, 2i+1)} = \cos(\text{pos} / 10000^{2i/d_{\text{model}}})
    \label{eq:pe_cos}
\end{equation}

\subsubsection{Parameter-Free Scaled Dot-Product Attention}
Next, we implement a simple parameter-free scaled self-attention mechanism. This is achieved by removing the learnable linear projection matrices typically used in transformer architectures to generate the Query (Q), Key (K), and Value (V) representations. The attention output is thus computed as:
\begin{equation}
    \text{Attention}(X_{pos}) = \text{softmax}\left(\frac{X_{pos} X_{pos}^T}{\sqrt{d_k}}\right)X_{pos}
    \label{eq:attention}
\end{equation}
where $d_k$ is the embedding dimension. This step aims to enrich the deep scattering representation with global context information, mimicking the traditional transformer architectures but without requiring backpropagation.
\subsubsection{Feed-Forward Network (FFN)}
This is then followed by a feed-forward network where random projection or feature selection can be used to limit the final embedding size as needed. Alternatively, this step can be omitted to preserve the context-aware scattering representation obtained.
\subsection{Operational Modes}
The proposed scattering transformer can be configured to operate in one of two modes depending on how the sequence fed to the transformer core module is constructed.
\begin{itemize}
    \item \textbf{Paths-as-Sequence Mode:} In this configuration, the standard scattering matrix (coefficients x paths) in Equation \ref{eq:scattering_matrix} is used as input to the further contextualization module. In essence, in this mode, the contextualization module extracts contextual information among different paths of the scattering matrix.
    \item \textbf{Multi-Segment Mode:} This configuration follows more the standard transformer architectures, where the contextualization is applied among various time segments of the signal. For each time segment, a scattering embedding is generated by averaging the coefficients in each path, resulting in an embedding with size $d_k$ equal to the number of paths. The input to the contextualization module is a matrix of (number of segments $\times$ $d_k$). This aims to capture long-term dependencies among different segments.
\end{itemize}

\subsection{Supervised Classification Head}
The output of the scattering transformer is context-aware scattering embeddings. This is then used in a supervised classification end task to detect heart murmur conditions. For this work, a Support Vector Machine (SVM) with a quadratic kernel was employed.
\section{Experiments and Results}\label{sec:experiments}
We benchmark the performance of the proposed Scattering Transformer on the heart murmur detection task from the George B. Moody PhysioNet Challenge 2022 \cite{reyna2023heart}. This allows for direct and fair comparison against state-of-the-art pre-trained foundation models as reported by Niizumi et. al \cite{niizumi2024exploring}.
\subsection{Experimental Setup}
The evaluation was performed on the three-class classification task (Present, Absent, Unknown) using phonocardiogram (PCG) recordings from the CirCor DigiScope dataset \cite{oliveira2022circor}.To prevent data leakage, we partitioned the data at the patient level into a training set (75\%) and a testing set (25\%). All audio recordings were resampled to 8 kHz and segmented into 5-second clips with a 2.5-second overlap. To mitigate the significant class imbalance inherent in the training dataset, we applied an oversampling strategy to the minority classes within the training set.

\subsection{Evaluation Protocol and Metrics}
The performance of the proposed method was evaluated using the official metrics from the George B. Moody PhysioNet Challenge 2022 to enable a direct fair comparison with state-of-the-art methods reported in \cite{niizumi2024exploring}. The metrics used in the evaluation are the Weighted Accuracy (W.acc) and the Unweighted Average Recall (UAR).

\subsubsection{The Weighted Accuracy (W.acc)}
This is the primary evaluation metric from the George B. Moody PhysioNet Challenge 2022. It assigns a greater penalty to the misclassifications of more clinically critical classes (Present, Unknown) and is defined in Equation \ref{eq:w_acc}.
\begin{equation}
    W.acc = \frac{5t_p + 3t_u + t_a}{5c_p + 3c_u + c_a}
    \label{eq:w_acc}
\end{equation}
where $c_i$ and $t_i$ represent the total number of true labels and the number of correct predictions for the present (p), unknown (u), and absent (a) classes, respectively.
\subsubsection{The Unweighted Average Recall (UAR)}
The Unweighted Average Recall (UAR) is a metric that reflects performance balance among classes; it is calculated as the macro-average of recall across all classes and provides an unbiased assessment of performance on an imbalanced dataset \cite{niizumi2024exploring}. It is defined as:
\begin{equation}
    \text{UAR} = \frac{1}{N_c} \sum_{i \in \{p, u, a\}} r_i
    \label{eq:uar}
\end{equation}
where $r_i$ is the recall for class $i$, and $N_c$ is the number of classes (three in this case).
\subsection{Patient-Level Aggregation}
A final diagnosis for each patient was determined by aggregating segment-level predictions. For this purpose, probability averaging was employed, whereby the raw prediction scores (logits) from the SVM classifier across all of a patient's audio segments were averaged. 
\subsection{Results: Comparison with Pre-trained Models}
Table \ref{tab:results} displays the performance of the proposed Scattering Transformer in comparison to a number of prominent pre-trained audio foundation models as reported in \cite{niizumi2024exploring}. Despite its lightweight and training-free architecture, our model attains a weighted accuracy (W.acc) of 0.786 and an unweighted average recall (UAR) of 0.697. These results position our model's performance with leading approaches such as wav2vec 2.0 as reported by Panah et al. \cite{panah2023exploring} and CUED\_Acoustics \cite{mcdonald2022detection} in the end task of automated murmur heart sound identification. Moreover, the Scattering Transformer markedly surpasses many other recognized deep learning models, such as AST \cite{gong2021ast}, BYOL-A \cite{niizumi2021byol}, and CNN14 \cite{kong2020panns}.
\begin{table}[h]
\caption{Comparison with Previous Methods on the PhysioNet 2022 Murmur Detection Task}\label{tab:results}
\centering
\begin{tabular}{@{}lccccc@{}}
\toprule
\textbf{Model} & \textbf{W.acc} & \textbf{UAR} & \textbf{Recall (P)} & \textbf{Recall (U)} & \textbf{Recall (A)} \\
\midrule
M2D (SOTA) \cite{niizumi2023masked} & 0.832 & 0.713 & 0.911 & 0.361 & 0.868 \\
Panah et al. \cite{panah2023exploring} & 0.800 & 0.70& 0.86& 0.410 & 0.830 \\
CUED\_Acoustics \cite{mcdonald2022detection} & 0.80& 0.68& 0.93& 0.340 & 0.780 \\
\textbf{Scattering Transformer (Ours)} & \textbf{0.786} & \textbf{0.697} & \textbf{0.800} & \textbf{0.692} & \textbf{0.600} \\
AST \cite{gong2021ast} & 0.654 & 0.672 & 0.744 & 0.769 & 0.505 \\
BYOL-A \cite{niizumi2021byol} & 0.556 & 0.556& 0.590& 0.573 & 0.507 \\
CNN14 \cite{kong2020panns} & 0.582 & 0.548 & 0.750& 0.506& 0.388\\
\bottomrule
\end{tabular}
\end{table}

\section{Discussion}\label{sec:discussion}
\subsection{Performance Profile and Class-Specific Strengths}
A detailed analysis of class-specific recall reveals a promising competitive performance profile for our proposed model compared to the top-achieving pre-trained audio foundation models. From Table \ref{tab:results}, while the M2D model \cite{niizumi2023masked} attains the best weighted accuracy, its capacity to recognize the "Unknown" class is significantly constrained, with a recall of 0.361. On the contrary, the Scattering Transformer achieves a recall of 0.692. This is nearly double that of the M2D model and significantly higher than Panah et al. \cite{panah2023exploring} (0.41). On the other hand, our model's recall for the "Absent" class (0.600) is lower than that of top performers. In terms of the UAR metric, our model achieves a competitive score of 0.697, indicating a balanced performance across all classes.
\subsection{The Critical Role of Contextualization: An Ablation Study}
\begin{figure}[h]
\centering
\begin{tikzpicture}
\begin{axis}[
    ybar,
    bar width=20pt,
    enlarge x limits=0.35,
    legend style={
        at={(0.05,0.95)},      
        anchor=north west,     
        legend columns=1,      
    },
    ylabel={Performance Metric Score},
    symbolic x coords={
        {WSN(Baseline)}, 
        {Scattering Transformer}
    },
    xtick=data,
    x tick label style={anchor=north},
    nodes near coords,
    nodes near coords align={vertical},
    ymin=0,
]
\addplot coordinates {
    ({WSN(Baseline)}, 0.481) 
    ({Scattering Transformer}, 0.786)
};

\addplot coordinates {
    ({WSN(Baseline)}, 0.464) 
    ({Scattering Transformer}, 0.697)
};

\legend{W.acc, UAR}
\end{axis}
\end{tikzpicture}
\caption{Ablation study comparing the performance of the proposed Scattering Transformer against the Standard Wavelet Scattering Networks (WSN) baseline. The chart displays the two primary evaluation metrics: Weighted Accuracy (W.acc) and Unweighted Average Recall (UAR).}
\label{fig:ablation_study_detailed}
\end{figure}
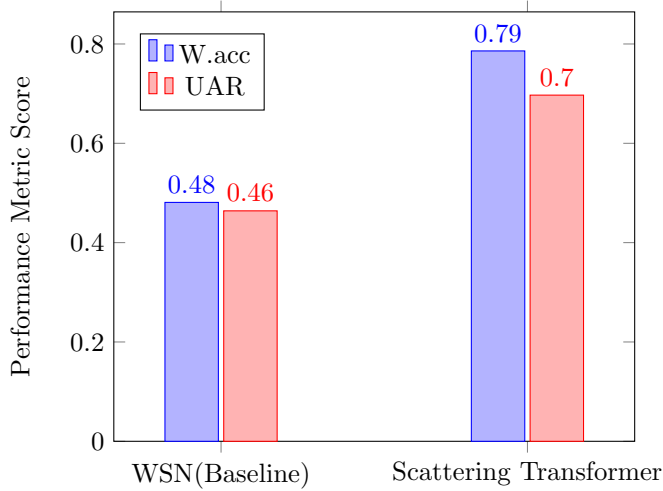

To quantify the contribution of our proposed parameter-free Scattering Transformer, we performed an ablation study by benchmarking a "Standard Scattering" model without the self-attention and contextualization mechanism. The results, illustrated in Figure \ref{fig:ablation_study_detailed}, underscore the critical importance of the contextualization step. The "Standard Scattering" baseline performed poorly in the end task of heart murmur sound identification, achieving a patient-level W.acc of only 0.481. On the other hand, the proposed model achieves a W.acc of 0.786, with a relative improvement of over 63\%. This provides clear evidence that our parameter-free attention mechanism is the crucial component that successfully injects global, sequential context into the local features, enabling powerful bio-signal representation and the end classification task of murmur heart sound identification.
\section{Conclusion}\label{sec:conclusion}
 This study introduced the Scattering Transformer as a context-aware self-attentive version of the standard wavelet scattering networks.
The Scattering Transformer, a lightweight and training-free transformer architecture, is an attractive alternative to computationally intensive foundation models for specific, data-scarce biomedical applications. Our evaluation on the problem of heart murmur detection using the CirCor DigiScope dataset validates this approach. The Scattering Transformer achieved a performance on par with state-of-the-art pre-trained audio foundation methods. This work, therefore, positions the Scattering Transformer as a viable and resource-efficient baseline for heart sound analysis, highlighting its potential as a valuable tool in other data-constrained biomedical contexts.
\bibliography{article}

\begin{thebibliography}{10}
\providecommand{\url}[1]{{#1}}
\providecommand{\urlprefix}{URL }
\providecommand{\doi}[1]{\url{https://doi.org/#1}}


\bibitem{who_cvd}
{World Health Organization}.
\newblock Cardiovascular diseases ({CVDs}).
\newblock Online (2025).
\newblock Accessed: 2025-08-30

\bibitem{ren2024comprehensive}
Z.~Ren, Y.~Chang, T.T. Nguyen, Y.T. Tan, K.~Qian, B.W. Schuller, A comprehensive survey on heart sound analysis in the deep learning era.
\newblock IEEE Computational Intelligence Magazine \textbf{19}(3), 42--57 (2024).
\newblock \doi{10.1109/mci.2024.3401309}

\bibitem{niizumi2024exploring}
D.~Niizumi, D.~Takeuchi, Y.~Ohishi, N.~Harada, K.~Kashino, Exploring pre-trained general-purpose audio representations for heart murmur detection.
\newblock arXiv preprint arXiv:2404.17107  (2024)

\bibitem{panah2023exploring}
D.S. Panah, A.~Hines, S.~McKeever, \emph{Exploring Wav2vec 2.0 Model for Heart Murmur Detection}, in \emph{2023 31st European Signal Processing Conference (EUSIPCO)} (2023), pp. 1010--1014.
\newblock \doi{10.23919/EUSIPCO58844.2023.10289947}

\bibitem{dwivedi2019algorithms}
A.K. Dwivedi, S.A. Imtiaz, E.~Rodriguez-Villegas, Algorithms for automatic analysis and classification of heart sounds-a systematic review.
\newblock IEEE Access \textbf{7}, 8316--8345 (2019).
\newblock \doi{10.1109/ACCESS.2018.2886457}

\bibitem{kong2020panns}
Q.~Kong, Y.~Cao, T.~Iqbal, Y.~Wang, W.~Wang, M.D. Plumbley, {PANNs: Large-Scale Pretrained Audio Neural Networks for Audio Pattern Recognition}.
\newblock IEEE/ACM Transactions on Audio, Speech, and Language Processing \textbf{28}, 2880--2894 (2020).
\newblock \doi{10.1109/TASLP.2020.3030497}

\bibitem{gong2021ast}
Y.~Gong, Y.A. Chung, J.~Glass, \emph{{AST: Audio Spectrogram Transformer}}, in \emph{Proc. Interspeech 2021} (2021).
\newblock \doi{10.21437/Interspeech.2021-698}

\bibitem{niizumi2023masked}
D.~Niizumi, D.~Takeuchi, Y.~Ohishi, N.~Harada, K.~Kashino, \emph{Masked Modeling Duo: learning representations by encouraging both networks to model the input}, in \emph{ICASSP 2023 - 2023 IEEE International Conference on Acoustics, Speech and Signal Processing (ICASSP)} (2023).
\newblock \doi{10.1109/icassp49357.2023.10097236}

\bibitem{oliveira2022circor}
J.H. Oliveira, F.~Renna, P.~Costa, M.~Nogueira, C.~Oliveira, C.~Ferreira, A.~Jorge, S.~Mattos, T.~Hatem, T.~Tavares, A.~Elola, A.B. Rad, R.~Sameni, G.D. Clifford, M.T. Coimbra, The {CirCor} {DigiScope} dataset: From murmur detection to murmur classification.
\newblock IEEE Journal of Biomedical and Health Informatics \textbf{26}(6), 2524--2535 (2022).
\newblock \doi{10.1109/JBHI.2021.3137048}

\bibitem{niizumi2021byol}
D.~Niizumi, D.~Takeuchi, Y.~Ohishi, N.~Harada, K.~Kashino, \emph{{BYOL for Audio: Self-Supervised Learning for General-Purpose Audio Representation}}, in \emph{2021 International Joint Conference on Neural Networks (IJCNN)} (2021).
\newblock \doi{10.1109/IJCNN52387.2021.9534474}

\bibitem{baevski2020wav2vec}
A.~Baevski, Y.~Zhou, A.r. Mohamed, M.~Auli, \emph{wav2vec 2.0: A framework for self-supervised learning of speech representations}, in \emph{Advances in Neural Information Processing Systems 33} (2020), pp. 12449--12460

\bibitem{gemmeke2017audio}
J.F. Gemmeke, D.P.W. Ellis, D.~Freedman, A.~Jansen, W.~Lawrence, R.C. Moore, M.~Plakal, M.~Ritter, \emph{Audio Set: An ontology and human-labeled dataset for audio events}, in \emph{2017 IEEE International Conference on Acoustics, Speech and Signal Processing (ICASSP)} (2017), pp. 776--780

\bibitem{reyna2023heart}
M.A. Reyna, Y.~Kiarashi, A.~Elola, J.~Oliveira, F.~Renna, A.~Gu, E.A. Perez~Alday, N.~Sadr, A.~Sharma, J.~Kpodonu, S.~Mattos, M.T. Coimbra, R.~Sameni, A.B. Rad, G.D. Clifford, Heart murmur detection from phonocardiogram recordings: The {G}eorge {B}. {M}oody {P}hysio{N}et {C}hallenge 2022.
\newblock PLOS Digital Health \textbf{2}(9), e0000324 (2023).
\newblock \doi{10.1371/journal.pdig.0000324}

\bibitem{koike2020audio}
T.~Koike, Y.~Bando, N.~Togawa, \emph{Audio for audio is better? an investigation on transfer learning models for heart sound classification} (2020), pp. 74--77

\bibitem{Han2025ENACTHeart}
J.~Han, A.~Shaout, {ENACT-Heart ENsemble-based Assessment Using CNN and Transformer on Heart Sounds}.
\newblock ArXiv \textbf{abs/2502.16914} (2025)

\bibitem{bruna2013invariant}
J.~Bruna, S.~Mallat, Invariant scattering convolution networks.
\newblock IEEE Transactions on Pattern Analysis and Machine Intelligence \textbf{35}(8), 1872--1886 (2013).
\newblock \doi{10.1109/TPAMI.2012.237}

\bibitem{hsdfenet}
H.~Chen, W.~Gu, A dual branch feature extraction network for heart sound signal analysis.
\newblock Scientific Reports \textbf{15}(1), 27557 (2025).
\newblock \doi{10.1038/s41598-025-12303-0}

\bibitem{mcdonald2022detection}
A.~McDonald, M.J. Gales, A.~Agarwal, \emph{Detection of Heart Murmurs in Phonocardiograms with Parallel Hidden Semi-Markov Models}, in \emph{2022 Computing in Cardiology (CinC)}, vol.~49 (2022), pp. 1--4

\end{thebibliography}

\end{document}